# Optimal scheduling of isolated microgrid with an electric vehicle battery swapping station in multi-stakeholder scenarios: a bi-level programming approach via real-time pricing

Yang Li [a,b,*], Zhen Yang [a], Guoqing Li [a], Yunfei Mu [c], Dongbo Zhao [b], Chen Chen [b], Bo Shen [d]

[a] School of Electrical Engineering, Northeast Electric Power University, Jilin 132012, China
[b] Energy Systems Division, Argonne National Laboratory, Lemont, IL 60439, USA
[c] Key Laboratory of Smart Grid of Ministry of Education, Tianjin University, Tianjin 300072, China
[d] Energy Analysis & Environmental Impacts Division, Lawrence Berkeley National Laboratory, Berkeley, CA 94720, USA

**Abstract**: In order to coordinate the scheduling problem between an isolated microgrid (IMG) and electric vehicle battery swapping stations (BSSs) in multi-stakeholder scenarios, a new bi-level optimal scheduling model is proposed for promoting the participation of BSSs in regulating the IMG economic operation. In this model, the upper-level sub-problem is formulated to minimize the IMG net costs, while the lower-level aims to maximize the profits of the BSS under real-time pricing environments determined by demand responses in the upper-level decision. To solve the model, a hybrid algorithm, called JAYA-BBA, is put forward by combining a real/integer-coded JAYA algorithm and the branch and bound algorithm (BBA), in which the JAYA and BBA are respectively employed to address the upper- and lower- level sub-problems, and the bi-level model is eventually solved through alternate iterations between the two levels. The simulation results on a microgrid test system verify the effectiveness and superiority of the presented approach.

## NOMENCLATURE

**Acronyms**

| | |
|---|---|
| MG | Microgrid |
| IMG | Isolated microgrid |
| WT | Wind turbine |
| PV | Photovoltaic |
| MT | Microturbine |
| EL | Equivalent load |
| BSS | Battery swapping station |
| EVs | Electric vehicles |
| DERs | Distributed energy resources |
| DGs | Distributed generations |
| JAYA | JAYA algorithm |
| BBA | Branch and Bound Algorithm |
| MILP | Mixed integer linear programming |
| BLP | Bi-level programming |
| SOT | Sequence operation theory |
| PDF | Probability density function |
| ATC | Addition-type-convolution |
| STC | Subtraction-type-convolution |
| HIA | Hybrid intelligent algorithm |

**Symbols**

| | |
|---|---|
| $q$ | Discrete step size (kW) |
| $F_1$ | Fuel cost of IMG ($) |
| $F_2$ | Total profit of BSS ($) |
| $T$ | Entire scheduling cycle (h) |
| $t$ | Time period (h) |
| $M_G$ | Total number of MT units |
| $P^{WT}$ | WT power output (kW) |
| $P^{PV}$ | PV power output (kW) |
| $P^L$ | Load active power (kW) |
| $P^{EL}$ | Power of the EL (kW) |
| $P_n^{MT}$ | MT output power (kW) |
| $R_n^{MT}$ | MT spinning reserve (kW) |
| $P_B^{DC}$ | BSS discharge power (kW) |
| $P_B^{CH}$ | BSS charge power (kW) |
| $P^{CNLAOD}$ | Controllable load output (kW) |
| $P_{B,Ress}$ | BSS reserve capacity (kW) |
| $N_{EV,t}$ | Number of EVs arriving during period $t$ |
| $\omega_{rt\_price,t}$ | Real-time price ($/kWh) |
| $\omega_{rc\_price}$ | Reserve price ($/kW) |
| $\omega_{rf\_price}$ | Reference price ($/kWh) |
| $\omega_{sp\_price}$ | Swap price ($/kWh) |
| $C_{B,min}$ | Minimal capacity of BSS (kWh) |
| $C_{B,max}$ | Maximum capacity of BSS (kWh) |
| $C_{B,0}$ | Initial capacity of BSS (kWh) |
| $N_{B,pos,max}$ | Maximum number of charge-discharge positions in BSS |
| $C_{bat,*}$ | Rated capacity of per battery (kWh) |
| $C_{B,*}$ | Rated capacity of BSS (kWh) |
| $\eta_B^{CH}$ | Charge coefficient of BSS (p.u) |
| $\eta_B^{CH}$ | Discharge coefficient of BSS (p.u) |
| $\alpha$ | Pre-given confidence level (%) |

**Subscripts**

| | |
|---|---|
| $w$ | Wind |
| $in$ | Cut-in |
| $out$ | Cut-out |
| $*$ | Rated value |
| $rf$ | Reference value |
| $r$ | Actual light intensities |
| $max$ | Maximum value |
| $min$ | Minimum value |
| $a$ | Probabilistic sequences of WT outputs |
| $b$ | Probabilistic sequences of PV outputs |
| $c$ | Probabilistic sequences of joint outputs |
| $d$ | Load probabilistic sequences |
| $e$ | Equivalent load probabilistic sequences |
| $n$ | Number of MT |
| $0$ | Initial energy |
| $Ress$ | Reserve capacity |
| $L$ | Load |

**Superscripts**

| | |
|---|---|
| $CH$ | Charge |
| $DC$ | Discharge |
| $CNLOAD$ | Controllable load |

* Corresponding author. E-mail address: liyang@neepu.edu.cn (Y. Li)

## 1. Introduction

As the energy crisis and environmental pollution become the major challenges of the automotive industry and urban management, research on the efficient use of distributed energy resources (DERs) and environment-friendly public transportation is very important to the sustainable development of countries and cities [1-4]. As a locally controlled system including interconnected loads and distributed generations (DGs), a microgrid (MG) is capable of connecting and disconnecting from the traditional centralized grid to enable it to flexibly and efficiently operate in both grid-connected or island-modes, which has proven to be useful for improving the receptivity of distribution system to DERs and enhancing the efficiency of renewable energy utilization [5, 6]. Compared with the grid-connected mode, an isolated MG (IMG) plays a unique role in guaranteeing the uninterruptible power supply of critical loads during extreme weather-related incidents such as thunderstorms, hurricanes, and blizzards [7]. In addition, it can also be used to supply power to users in remote areas that are inaccessible to the main power grid involving rural villages, islands, and deserts [8, 9]. On the other hand, electric vehicles (EVs), as an emerging mode of transportation, are being increasingly accepted and adopted not only because they are helpful to reduce dependence on fossil fuels, but also because they emit less greenhouse gas (GHG) emissions than traditional diesel-engine vehicles [10, 11]. However, there are still some open problems in the promotion of EVs such as long battery charging times, limited EV range and high battery replacement cost [12], which, to a certain extent, limits the usefulness and popularity of EVs. For addressing these issues, a battery swapping station (BSS) as an important way to implement vehicle-to-grid (V2G) technology offers an effective solution and has received ever-increasing attention in the past few years [12]. As a battery aggregator, BSSs are capable of participating in markets for electrical energy and reserve. The BSS can achieve its maximum profits by supplying services to the system, such as voltage support, regulation reserves, energy arbitrage, or black-start support [13]. With the growing maturity of the battery storage technology, the utilization of BSSs has been moving toward large-scale commercialization. For example, Sun Mobility is developing a service for swapping electric bus batteries, as well as smaller two- and three-wheel vehicles, in India. In 2013, Tesla Motors introduced battery swapping technology for their EVs [12]. From the perspective of power grids, BSS is a flexible and uncertain load demand; while in the eyes of EV users, it is a service supplier that provides them with fully charged batteries with charging a certain amount of fee. Furthermore, recent research shows that the integration of BSSs into an MG is a 'win-win' strategy for both of them, which is an important development trend of future smart grids [14].

### 1.1 Literature review

There has been a significant amount of studies carried out on different aspects of MG and BSS. First, research on MGs has been undergoing a boom around the world for the last past years after demonstrating its great value in critical situations with a very wide scope covering various aspects such as planning, operation, and control strategies [5-9, 15, 16 17]. Regarding BSS, there have already been pioneer works for resolving the issues about its planning and operation. An optimization framework for the BSS operating model is proposed under consideration of the battery demand uncertainty and day-ahead scheduling process in reference [12]. Reference [18] presents an automatic generation control strategy with the use of controllable energy storage in BSSs for frequency regulation. Reference [19] puts forward a centralized charging strategy and dispatch algorithm of EVs under the battery swapping scenario by taking into account optimal charging priority and charging location. To seek the minimum cost, an optimization model is proposed for EV charging at a BSS in reference [20]. To analyze and optimize battery swapping behaviors of EV users, an optimization charging mode is presented based on an actual survey in reference [21]. Reference [22] proposes a charging strategy for a photovoltaic (PV)-based BSS under consideration of the service availability and PV energy self-consumption. An optimal design approach is proposed for BSSs in distribution systems using life-cycle cost in reference [23]. In reference [24], optimal charging schedule for a BSS serving electric bus has been investigated.

Unfortunately, up to the present time, there are only quite a few works that have concerned the scheduling problem of an MG incorporating BSSs in literature [25, 26]. In reference [25], a business model of BSS is put forward and an optimal scheduling strategy of MG with BSSs is given as well. Through the use of interruptible-loads dispatch, a strategy for energy management and optimal operation model of EVs is proposed in reference [26] to minimize operating costs and maximize profits of an MG. However, all these reported pioneering works assume that the BSS charge-discharge scheme, including charge-discharge state and trading power, is fully controlled by an MG control center without consideration of the interest demands of the BSS. But in fact, the BSS, as an independent stakeholder, has multiple operation modes and incentive measures is an important adjustment means to guide the BSS charge and discharge behavior such that the interests between MG and BSS can be better coordinated in multi-stakeholder scenarios [27, 28]. Meanwhile, due to relatively small capacity and inherently intermittent nature of DGs, the operation of IMG is vulnerable to uncertain power exchange between the sources and the loads, and thus it is a very challenging task to handling economic dispatching of IMG with BSS while guaranteeing its operational reliability and supply security [29]. To the best of authors' knowledge, until now no study in the literature has yet comprehensively investigated the coordinated scheduling problem between IMG and BSS in multi-stakeholders scenarios. Taking into account the obvious hierarchical features between IMG and BSS, a bi-level programming model via real-time pricing is proposed to achieve the optimal day-ahead scheduling of IMG with BSS. In general, it is very tricky to resolve a bi-level programming problem, since it is known to be non-deterministic polynomial-time hard (NP-hard). To solve this model, a hybrid approach combining the heuristic with analytical optimization, called JAYA-branch and bound algorithm (JAYA-BBA), is thereby developed in this work.

**1.2 Contribution of This Paper**

The main contributions are highlighted as follows:

- A new MG scheduling model — to coordinate the scheduling between IMG and BSS, a new MG optimal day-ahead scheduling model based on bi-level programming is proposed in multi-stakeholder scenarios for promoting participation of BSSs in regulating the IMG economic operation. In this model, a new real-time pricing mechanism based on demand responses is designed under consideration of dynamic supply-demand relationships between IMG and BSS.
- A novel solution methodology — a hybrid approach JAYA-BBA is developed by combining the heuristic with analytical optimization, which uses JAYA to obtain IMG scheduling schemes by solving the upper level, and the lower level gives the BSS charge-discharge scheduling scheme using BBA under real-time pricing environments determined in the upper-level decision; eventually, the bi-level model is solved through alternate iterations between the two levels.
- The proposed approach is examined against a modified Oak Ridge National Laboratory (ORNL) Distributed Energy Control and Communication (DECC) lab MG test system. The simulation results verify the effectiveness and superiority of the presented approach.

**1.3 Organization of This Paper**

The rest of this paper is organized as follows. An introduction of uncertainty modeling of IMG with BSS is provided in Section 2; and then, Section 3 gives the problem formulation in detail; in the next place, its further solution methodology is displayed in Section 4. Case studies are performed in Section 5; and finally, the conclusions are drawn in Section 6.

## 2. The model of MG

**2.1 Probabilistic WT Model**

The uncertainty of WT outputs is mainly originated from the inherent intermittency of wind speeds. Previous research demonstrates that wind speed follows the Weibull distribution [30, 31]. The probability density function (PDF) of wind speed is accordingly given by [31]

$$f_w(v) = (k/\gamma)(v/\gamma)^{k-1} \exp[-(v/\gamma)^k] \tag{1}$$

where $v$ represents the actual wind speed; $k$ is the shape factor (dimensionless), which describe the PDF shape of wind speed; $\gamma$ is the scale factor.

Since the locations of WT are geographically close in an MG, it is shown that correlations of wind speeds in different places should be taken into account for the probabilistic analysis of system performances [31]. The correlation coefficient with a range [−1, 1] is an effective indicator to assess how well wind speeds at two sites follow each other.

The relationship between the WT power output $P^{WT}$ and the actual wind speed $v$ can be described as [31]:

$$P^{WT}(v) = \begin{cases} 0 & v < v_{in}, v > v_{out} \\ \dfrac{v - v_{in}}{v_* - v_{in}} P_*^{WT} & v_{in} \leq v < v_* \\ P_*^{WT} & v_* \leq v < v_{out} \end{cases} \tag{2}$$

where $P_*$ denotes the rated output power of a WT, $v_{in}$ is the cut-in wind speed, $v_{out}$ is the cut-out wind speed, and $v_*$ is the rated wind speed.

According to Eq. (1) and (2), the PDF of the WT output $f_o(P^{WT})$ can be formulated as

$$f_o(P^{WT}) = \begin{cases} (khv_{in}/\gamma P_*)\left[((1+hP^{WT}/P_*)v_{in})/\gamma\right]^{k-1} \times \exp\left\{-\left[((1+hP^{WT}/P_*)v_{in})/\gamma\right]^k\right\}, p^{WT} \in [0, P_*] \\ 0, \quad \text{otherwise} \end{cases} \tag{3}$$

where $h = (v_*/v_{in}) - 1$.

**2.2 Probabilistic PV Model**

The PV output is mainly dependent on the amount of solar irradiance reaching the ground, ambient temperature and characteristics of the PV module itself. The statistical study shows that the solar irradiance for each hour of the day follows the Beta distribution [29], which is a set of continuous probability distribution functions defined in the interval (0, 1). The Beta PDF used to depict the probabilistic nature of solar irradiance is

$$f_r(r) = \frac{\Gamma(\lambda_1) + \Gamma(\lambda_2)}{\Gamma(\lambda_1)\Gamma(\lambda_2)} \left(\frac{r}{r_{max}}\right)^{\lambda_1 - 1} \left(1 - \frac{r}{r_{max}}\right)^{\lambda_2 - 1} \tag{4}$$

where $r$ and $r_{max}$ are respectively the actual and maximum light intensities; $\lambda_1$ and $\lambda_2$ are the shape factors; $\Gamma$ represents a Gamma function in the following form: $\Gamma(\lambda) = \int_0^{+\infty} \rho^{\lambda-1} e^{-\rho} d\rho$, wherein $\rho$ is an integration variable. The relationship between PV outputs and solar irradiances used in this study is [30]

$$P^{PV} = \xi A^{PV} \eta^{PV} \tag{5}$$

where $\xi$ is the solar irradiance, $A^{PV}$ is the radiation area of this PV, and $\eta^{PV}$ is the conversion efficiency.

From Eq. (5) it can be seen that the PV output is linear with the solar irradiance, and thereby, the PV output is also generally subject to the Beta distribution. The PDF of PV output is given as follows [30]:

$$f_p(P^{PV}) = \frac{\Gamma(\lambda_1)+\Gamma(\lambda_2)}{\Gamma(\lambda_1)\Gamma(\lambda_2)}\left(\frac{P^{PV}}{P_{max}^{PV}}\right)^{\lambda_1-1}\left(1-\frac{P^{PV}}{P_{max}^{PV}}\right)^{\lambda_2-1} \tag{6}$$

where $P^{PV}$ and $P_{max}^{PV}$ represent the output of this PV and its maximum value, respectively.

### 2.3 Probabilistic load injection model

A widely used normal distribution model is adopted here for modeling load fluctuations. The PDF can be described as [31]

$$f_l(P^L) = \frac{1}{\sqrt{2\pi}\sigma_L}\exp\left(-\frac{(P^L-\mu_L)^2}{2\sigma_L^2}\right) \tag{7}$$

where $P^L$ is the load active power, $\mu_L$ and $\sigma_L$ are the mean and standard deviation of $P^L$.

### 2.4 Equivalent Load Model

In order to facilitate the incorporation of multiple random variables, the power of an equivalent load (EL) is defined as the difference of the load power and the joint power output of WT and PV, which is expressed as follows:

$$P^{EL} = P^L - (P^{WT}+P^{PV}) \tag{8}$$

where $P^{EL}$ denotes the power of the EL.

### 2.5 BSS operations

As a mediator between the MG and EV users, BSS provides battery swap service for their customers on request a fee. Once the energy stored in batteries of EVs is not sufficient for the next trip, EVs users have the option to swap it with the one fully charged in BSS. The empty battery will be recharged in BSS. The fully charged ones are prepared for the battery swap service and provide reserve capacity to system together. For ease of analysis, four basic assumptions are made in this paper. (1) The battery in charge or discharge does not provide any other battery swap services, and the BSS can be either charged or discharged at the same time [32]. (2) The battery swap time is not considered in the model and the empty batteries are fed at next time period [33]. (3) The internal battery capacity of BSS is consistent, and the battery charged and discharged with constant power, as well as the power conversion efficiency of the battery is 95% [11]. (4) To facilitate the subsequent analysis without loss of generality, we assume that an EV user can complete the battery swap operation by replacing only one battery at a time.

Fig. 1 shows the key features of BSS operations. The BSS maximizes its profits in the following way: first, purchase electricity from the IMG and charge batteries in Grid-to-Battery (G2B) mode during the low-priced time periods of the day; and then, sell the energy stored in batteries by discharging batteries to the IMG in Battery-to-Grid (B2G) mode during the high-price periods [12].

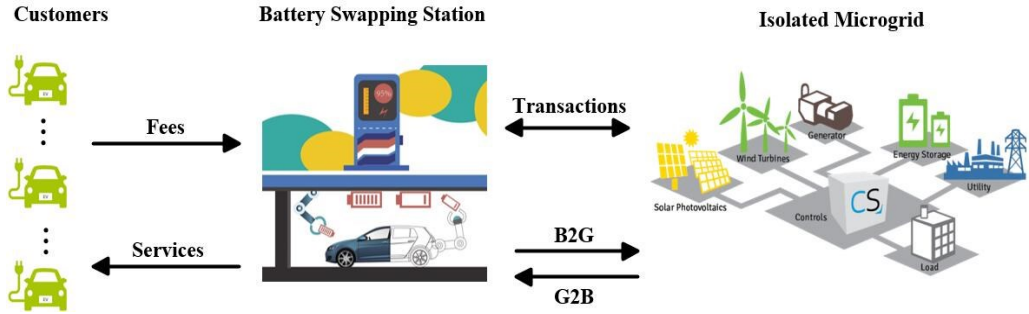

**Fig. 1.** BSS interactions with customers, market, and the IMG

### 2.6 Arrival time of EVs

A Poisson distribution is a well-known discrete distribution which is suitable for modeling the occurrence of waiting events. Recent studies have demonstrated that the number of EVs arriving at the BSS in a time period $t$ is considered to follow the Poisson distribution [34, 35]. The probability of $n_{ev}$ vehicles arriving during period $t$ is given by

$$Prb\{N_{ev}(t)=n_{ev}\} = \frac{e^{-\lambda_{ev}t}(\lambda_{ev}t)^{n_{ev}}}{n_{ev}!}, \quad n_{ev}=0,1,\cdots \tag{9}$$

where $Prb\{\cdot\}$ denotes the probability of an event, $N_{ev}(t)$ is the number of EVs arriving at the BSS in a time period $t$, the parameter $\lambda_{ev}$ is the arrival rate of the Poisson process which determines the frequency of EVs arriving at the BSS.

## 3. Problem formulation

The proposed bi-level model has two components, i.e., an upper-level sub-problem to model the operation of the IMG and a lower-level sub-problem to formulate the operation of the BSS under real-time pricing environments determined by demand responses [37-39]. It should be noted that the upper-level objective function is constrained not only by the upper-level constraints but also by the lower-level sun-problem. The upper and lower-level sub-problems are first discussed, and then the proposed bi-level model is transformed into a single-level mathematical program with equilibrium constraints (MPEC) model and solved thereafter.

## 3.1 Real-time-pricing mechanism

Considering dynamic electricity supply-demand relationships, a real-time pricing mechanism based on demand responses is proposed according to the following two principles:

***Rule #1***: when the BSS buys/sells electricity from/for the IMG, the electricity price depends on the overall load level of the system (defined as the algebraic sum of the IMG equivalent load and the electrical quantity traded between BSS and IMG). If the overall load level is greater than the IMG equivalent load, the price should be greater than a reference price; otherwise, the price should be lower than the reference price.

***Rule #2***: When the BSS only provides spinning reserve ancillary services to participate in the operation regulation of the IMG, the IMG should pay the BSS certain fees. In such cases, the BSS doesn't buy/buy electricity from/to IMG.

The proposed pricing mechanism is described in detail as follows.

(1) First, obtain the equal load at period $t$ according to the sequence operation theory (SOT).

(2) And then, the BSS charge-discharge scheme is obtained through the lower-level optimization and it will be feedback to the upper layer.

(3) The upper-level determines the real-time electricity price according to the following formula:

$$\omega_{rt\_price,t} = \begin{cases} \dfrac{P_t^{EL} \times \Delta t + S_{B,t}}{P_{rf}^{EL} \times \Delta t} \times \omega_{rf\_price} & U_{B,t} = 1 \\ \dfrac{P_t^{EL}}{P_{rf}^{EL}} \times \omega_{rf\_price} & U_{B,t} = 0 \end{cases} \tag{10}$$

where $\omega_{rt\_price,t}$ is the real-time price between BSS and IMG; and $\omega_{rf\_price}$ is a constant number, which denotes the reference price; $P_{rf}^{EL}$ denotes the pre-defined reference power of EL; $\Delta t$ is the duration of a time period, which is here taken as 1 hour; $S_{B,t}$ is the electrical quantity traded between BSS and IMG, which is determined by the BSS charge-discharge scheme, if it is positive representing BSS is in the charging state, on the contrary, it is negative number which denotes the BSS is in the discharging state. $U_{B,t}$ is a 0-1 state variable representing the operating state of the BSS: if the BSS is in the charge-discharge state, $U_{B,t}$ is set to 1; otherwise, the BSS is in the state that provides reserve capacity for IMG, it is 0.

## 3.2 Upper-level sub-problem
### 3.2.1 Objective function

The MG scheduling problem can be formulated as a nonlinear finite-horizon optimal control problem with multiple random variables. Due to the uncontrollability of renewable generations, the total IMG net cost comprises the following four folds: the charge-discharge cost, the spinning reserve cost, the fuel cost of MT units, and the cost of reserve provided by BSS. Thus, the objective function of the upper-level model is formulated to minimize the IMG net cost $F_1$, which is given as follows:

$$\min F_1 = -\sum_{t=1}^{T}(\omega_{rt\_price,t} \times S_{B,t} \times U_{B,t}) + \sum_{t=1}^{T}\sum_{n=1}^{M_G}\varsigma_n R_{n,t}^{MT} + \sum_{t=1}^{T}[\sum_{n=1}^{M_G}(\kappa_n S_{n,t} + U_{n,t}(\zeta_n + \psi_n P_{n,t}^{MT}))] + \omega_{rc\_price} \times \sum_{t=1}^{T}(1-U_{B,t}) \times P_{B,Ress,t} \tag{11}$$

where $t$ is the scheduling time period in hours, $T$ is the total number of time periods in a scheduling cycle (here $T$ is taken as 24), $n$ is the MT number, $M_G$ is the total number of MT units, $\zeta_n$ and $\psi_n$ are the MT consumption factors, $U_{n,t}$ and $S_{n,t}$ are 0-1 variables representing respectively the state variables and start-up variables, $\varsigma_n$ and $\kappa_n$ are respectively the spinning reserve costs and the start-up costs, $P_{n,t}^{MT}$ and $R_{n,t}^{MT}$ represent the MT output power and spinning reserve of during period $t$, respectively. $\omega_{rc\_price}$ is the reserve price for the IMG, $P_{B,Ress,t}$ is the reserve provided by BSS for IMG during period $t$.

### 3.2.2 Constraints
#### 3.2.2.1 System power balance constraint

In order to avoid an oversupply of renewable DGs, it is necessary to equip controllable loads for maintaining the power balance in an IMG when the DG outputs are too large. Consequently, the power balance constraint can be expressed as

$$\sum_{n=1}^{M_G} P_{n,t}^{MT} + P_{B,t}^{DC} - P_{B,t}^{CH} = E(P_t^{EL}) + P_t^{CNLOAD}, \forall t \tag{12}$$

where $P_{B,t}^{CH}$ and $P_{B,t}^{DC}$ are respectively the charge and discharge powers of BSS during period $t$, and they both are collectively called exchanging powers. $P_t^{CNLOAD}$ is the output of the controllable load, $P_t^{EL}$ is the predicted value of the EL power, and $E(P_t^{EL})$ is the expected value of $P_t^{EL}$ which is given by

$$E(P_t^{EL}) = \sum_{u_{d,t}=0}^{N_{d,t}} u_{d,t} qa(u_{d,t}) - \sum_{u_{a,t}=0}^{N_{a,t}} u_{a,t} qa(u_{a,t}) - \sum_{u_{b,t}=0}^{N_{b,t}} u_{b,t} qa(u_{b,t}) \tag{13}$$

#### 3.2.2.2 MT output constraint

In order to ensure the power output of an MT unit within the allowable range, it must obey the following inequality:

$$U_{n,t} P_{n,min}^{MT} \leq P_{n,t}^{MT} \leq U_{n,t} P_{n,max}^{MT}, \forall t, n \in M_G \tag{14}$$

where $P_{n,min}^{MT}$ and $P_{n,max}^{MT}$ represent the minimal and maximum outputs of the $n$-th MT unit, respectively.

#### 3.2.2.3 Spinning reserve constraints

For an IMG, due to unavailability of power support from main grids, the spinning reserve is the most important resource for leveling off the fluctuating power outputs of intermittent DGs and ensuring reliable operation of the system [30]. In this study, the required spinning reserves are provided by the MT units. Correspondingly, the spinning reserve constraint is

$$P_{n,t}^{MT} + R_{n,t}^{MT} \leq U_{n,t} P_{n,max}^{MT}, \forall t, n \in M_G \quad (15)$$

Eq. (12) shows that the MG overall uncertainty including source and load sides is mainly reflected in the equivalent load. Hence, to maintain the power balance, the total spinning reserve provided by MTs is used to compensate for the difference between the fluctuating EL power and its expected value.

Generally, the operation of IMG is susceptible to uncertainties from both sources and loads due to its relatively small capacity and unavailability of power supports from the main grid. In terms of the IMG studied in this work, there are multiple DGs with different probability distribution characteristics on the source side, while as far as the load side the swap demand of the BSS increase the uncertainty together with the original load. And thereby, sufficient spinning reserves are required to maintain the system reliable operation, this will inevitably lead to a very high reserve cost. For this reason, it is no doubt a preferable choice to model the spinning reserve requirement as a probability constraint to balance reliability and economy.

$$Prb\left\{\sum_{n=1}^{M_G} R_{n,t}^{MT} + P_{B,Ress,t} \geq (P_t^L - P_t^{WT} - P_t^{PV}) - E(P_t^{EL})\right\} \geq \alpha, \forall t \quad (16)$$

where $\alpha$ is the pre-given confidence level.

### 3.3 Lower-level sub-problem
#### 3.3.1 Objective function

The BSS total profit comprises the following four aspects: charge-discharge costs, swapping incomes, BSS depreciation costs, and reserve incomes. And thereby, the objective function of the lower-level model is modeled to maximize the BSS profits $F_2$, which is as follows:

$$\max F_2 = -\sum_{t=1}^{T}(\omega_{rt\_price,t} \times S_{B,t} \times U_{B,t}) + \omega_{sp\_price} \times C_{bat,*} \times \sum_{t=1}^{T} N_{EV,t} - \tau \sum_{t=1}^{T}\left(N_{EV,t} + P_{B,t}^{DC} \times \Delta t / C_{bat,*}\right) + \omega_{rc\_price} \times \sum_{t=1}^{T}(1 - U_{B,t}) \times P_{B,Ress,t} \quad (17)$$

where $\tau$ is the depreciation coefficient of a charge-discharge process for a battery, $C_{bat,*}$ represents the rated capacity of per battery in the BSS, $N_{EV,t}$ is the number of EVs arriving at BSS during period $t$, $\omega_{sp\_price}$ is the swap price provided by BSS to EV users.

#### 3.3.2 Constraints

BSS not only provides rapid battery swap service, but also participate in the regulation of IMG via real-time price between IMG and BSS, and it has been successfully utilized in MGs to balance random fluctuations, maintain system reliability and improve power quality [6, 35].

##### 3.3.2.1 BSS constraints

**Charge-discharge equation**: This constraint depicts the BSS charge-discharge states. The relationship between the available capacity of BSS during period $t+1$ and the charging-discharging powers during period $t$ is expressed as

$$\begin{cases} C_{B,t+1} = C_{B,t} + \eta_B^{CH} P_{B,t}^{CH} \Delta t \\ C_{B,t+1} = C_{B,t} - \Delta t P_{B,t}^{DC} / \eta_B^{DC} \end{cases}, \forall t \quad (18)$$

where $C_{B,t+1}$ and $C_{B,t}$ are the available capacity of the BSS during period $t+1$ and $t$, $\eta_B^{CH}$ and $\eta_B^{DC}$ are respectively the charge and discharge efficiencies of the BSS.

**Charge-discharge rate limits**: In order to ensure the BSS charge-discharge rate within the allowable range, the charge-discharge rate of BSS must obey the following constraints [26]:

$$\begin{cases} 0 \leq P_{B,t}^{CH} \leq P_{B,*}^{CH} \\ -P_{B,*}^{DC} \leq P_{B,t}^{DC} \leq 0 \end{cases} \forall t \quad (19)$$

where $P_{B,*}^{CH}$ and $P_{B,*}^{DC}$ are respectively the rated charge and discharge powers of the BSS.

**Capacity constraint**: To limit the BSS capacity during each period to a reasonable range, it must obey the following constraint [26, 40]:

$$C_{B,min} \leq C_{B,t} + \eta_B^{CH} P_{B,t}^{CH} + \frac{P_{B,t}^{DC}}{\eta_B^{DC}} \leq C_{B,max} \quad (20)$$

where $C_{B,max}$ and $C_{B,min}$ are the maximum and the minimal capacities of the BSS, respectively. The minimum capacity $C_{B,min}$ of BSS is assumed to be 20% of its own total capacity [41], i.e. the BSS doesn't work if the remaining capacity of BSS is lower than the $C_{B,min}$ value.

**Expected charging power constraint**: The charging power of BSS must obey the following constraint [26]:

$$C_{B,min} \leq \sum_{t=1}^{T}\left(\eta_B^{CH} P_{B,t}^{CH} + P_{B,t}^{DC} / \eta_B^{DC}\right) + C_{B,0} \leq C_{B,max} \quad (21)$$

where $C_{B,0}$ is the initial capacity of the BSS. This is to ensure that the capacity of BSS will remain within an allowable range in an entire scheduling cycle.

**Reserve capacity constraint**: In order to make the BSS reserve capacity not exceed its available capacity, the required spinning reserves of the system are also provided by the BSS considering its ability to participate in ancillary services.

$$P_{B,Ress,t} \leq \eta_B^{DC}(C_{B,t} - C_{B,\min})/\Delta t, \quad \forall t \in T_{Ress} \tag{22}$$

where $T_{Ress}$ denotes all time periods that BSS provides reserve capacity for IMG.

### 3.3.2.2 Battery constraints

**Total number of batteries balance constraint**: The total number of batteries in BSS must obey the following constraint:

$$N_{bat,ful,t} + N_{bat,emp,t} + N_{bat,t}^{CH} + N_{bat,t}^{DC} = N_{bat}, \quad \forall t \tag{23}$$

where $N_{bat,ful,t}$, $N_{bat,emp,t}$ are respectively the number of batteries with a full charge and empty charge during period $t$; $N_{bat,t}^{CH}$ and $N_{bat,t}^{DC}$ are the number of batteries with charging and discharging; $N_{bat}$ represents the total number of batteries in the BSS.

**Charge-discharge quantity of battery constraint**: In order to ensure that the number of batteries in the state of charge-discharge does not exceed the number of charge-discharge positions in the BSS, the following constraint should be satisfied

$$N_{bat,t}^{CH} \leq N_{B,pos,\max}, \quad N_{bat,t}^{DC} \leq N_{B,pos,\max} \tag{24}$$

where $N_{B,pos,\max}$ is the maximum number of charge-discharge positions in BSS.

**Charge-discharge state constraint:**
The constraint about the state of charge-discharge of batteries is

$$C_{bat,t+1} = C_{bat,t} + (\eta_{ch} P_{bat,t}^{CH} - P_{bat,t}^{DC}/\eta_{dc})\Delta t, \forall t \tag{25}$$

where $C_{bat,t+1}$ and $C_{bat,t}$ are the battery capacity in BSS until period $t+1$ and $t$, $\eta_{ch}$ and $\eta_{dc}$ are respectively the charge-discharge efficiencies of batteries. $\Delta t$ is the duration of a time period, which is here taken as 1 hour. $P_{bat}^{CH}$ and $P_{bat}^{DC}$ are the charge and discharge powers of the battery during period $t$, respectively.

**Capacity constraint:**
In order to ensure the BSS capacity within the allowable range, the capacity of BSS must obey the following constraints

$$C_{bat,\min} \leq C_{bat,t} \leq C_{bat,\max}, \forall t \tag{26}$$

where $C_{bat,\min}$ and $C_{bat,\max}$ are the minimum and maximum capacity of the battery during period $t$, respectively.

**Charge-discharge cycle constraint:**
Recent investigations show that it's a critical problem to limit the number of charge/discharge cycles in the design and operation of a battery energy storage system, since frequent charge/discharge cycles might seriously damage the health of batteries by shortening the lifetime of batteries [42, 43]. In this respect, the number of charge/discharge cycles for each battery is limited within a reasonable range, which is given as follows:

$$N_{bat,cyc} \leq N_{bat,cyc,\max} \tag{27}$$

where $N_{bat,cyc}$ is the battery charge-discharge cycles in an entire microgrid scheduling cycle, $N_{bat,cyc,\max}$ is the pre-given maximum value of $N_{bat,cyc}$.

## 4. Solution methodology

Taking into account the proposed bi-level programming-based model is NP-hard, an acceptable solution should satisfy the two criteria: the sufficient optimality and good computational efficiency. For this purpose, the hybrid approach JAYA-BBA is developed to solve the proposed model, which respectively use JAYA and BBA to address the upper- and lower- level sub-problems; the upper and lower levels are only connected by the real-time price and BSS charge-discharge schemes. In this way, this hybrid enables the proposed model to be solved with sufficient optimality and good computational efficiency by combing the advantages of heuristic and analytical approaches.

In this section, the proposed hybrid approach JAYA-BBA for solving the built bi-level scheduling model is described in detail. First, the serialization description of random variables is introduced; and then, the transformation of chance constraints into their deterministic form is described; next, the basic principles of the Jaya algorithm and BBA are briefly presented; what's more, how to calculate the optimal scheme is given, and finally, the specific solving process is listed.

### 4.1 Serialization description of random variables

#### 4.1.1 Introduction of SOT

SOT is a powerful mathematical tool for handling the uncertainties of power systems, which is based on the sequence convolution in the field of digital signal processing [36]. According to this theory, the probability distributions of random variables are denoted by probabilistic sequences, and a new sequence could be derived from the operations between sequences. And then, the new probability distributions of random variables are obtained through mutual operations [36]. Specifically, four discrete sequence operations named addition-type-convolution (ATC) and subtraction-type-convolution (STC), AND-type-product and OR-type-product have been defined.

**Definition 1.** Suppose a discrete sequence $a(i)$ with the length $N_a$, $a(i)$ is called a probabilistic sequence if

$$\sum_{i=0}^{N_a} a(i) = 1, \ a(i) \geq 0, \ i = 0,1,2,...,N_a \tag{28}$$

**Definition 2.** Given a probabilistic sequence $a(i)$ with the length $N_a$, its expected value is defined as follows:

$$E(a) = \sum_{i=0}^{N_a}[i\,a(i)] = \sum_{i=1}^{N_a}[i\,a(i)] \tag{29}$$

### 4.1.2 Sequence description of intermittent DG outputs

During period $t$, the WT output $P_t^{WT}$, the PV output $P_t^{PV}$, and the load power $P_t^L$ are random variables. In this theory, all these random variables are depicted by the corresponding probabilistic sequences $a(i_{a,t})$, $b(i_{b,t})$ and $c(i_{c,t})$ through discretization of continuous probability distributions. The length of the WT output probabilistic sequence $N_{a,t}$ is calculated by using the following equation:

$$N_{a,t} = [P_{\max,t}^{WT}/q] \tag{30}$$

where $q$ denotes the discrete step size and $P_{\max,t}^{WT}$ is the maximum value of WT power output during period $t$.

Table 1 shows the WT outputs and their corresponding probabilistic sequences.

**Table 1** WT outputs and their probabilistic sequence

| Power (kW)  | 0    | $q$  | … | $u_a q$  | … | $N_{a,t}q$   |
|---|---|---|---|---|---|---|
| Probability | $a(0)$ | $a(1)$ | … | $a(u_a)$ | … | $a(N_{a,t})$ |

The probabilistic sequence of the WT output can be calculated by using the PDF of the WT output:

$$a(i_{a,t}) = \begin{cases} \int_0^{q/2} f_o(P^{WT})dP^{WT}, & i_{a,t}=0 \\ \int_{i_{a,t}q-q/2}^{i_{a,t}q+q/2} f_o(P^{WT})dP^{WT}, & i_{a,t}>0, i_{a,t}\neq N_{a,t} \\ \int_{i_{a,t}q-q/2}^{i_{a,t}q} f_o(P^{WT})dP^{WT}, & i_{a,t}=N_{a,t} \end{cases} \tag{31}$$

## 4.2 Treatment of chance constraints

### 4.2.1 Probabilistic sequence of the equivalent load power

It is assumed that uncertainties of WT, PV and loads are independent, which is a common assumption. Given that the probabilistic sequences of $P_t^{PV}$ and $P_t^{WT}$ are respectively denoted as $a(i_{a,t})$ and $b(i_{b,t})$, the probabilistic sequence $c(i_{c,t})$ of their joint power outputs is obtained by the ATC operation in SOT theory [36], i.e.

$$c(i_{c,t}) = a(i_{a,t}) \oplus b(i_{b,t}) = \sum_{i_{a,t}+i_{b,t}=i_{c,t}} a(i_{a,t})b(i_{b,t}),\ i_{c,t}=0,1,...,N_{a,t}+N_{b,t} \tag{32}$$

Given the probabilistic sequence of $P_t^L$ is $d(i_{d,t})$ with length $N_{d,t}$, the probabilistic sequence of EL power $e(i_{e,t})$ is calculated by the STC operation [36]:

$$e(i_{e,t}) = d(i_{d,t}) \ominus \begin{cases} \sum_{i_{d,t}-i_{c,t}=i_{e,t}} d(i_{d,t})c(i_{c,t}), & 1 \leq i_{e,t} \leq N_{e,t} \\ \sum_{i_{d,t}\leq i_{c,t}} d(i_{d,t})c(i_{c,t}), & i_{e,t}=0 \end{cases} \tag{33}$$

The corresponding relation between the probabilistic sequence of the equivalent load power $P^{EL}$ with the step size $q$ and the length $N_{e,t}$ is shown in Table 2.

**Table 2** Probabilistic sequence of EL power

| Power (kW)  | 0    | $q$  | … | $u_e q$  | … | $N_{e,t}q$   |
|---|---|---|---|---|---|---|
| Probability | $e(0)$ | $e(1)$ | … | $e(u_e)$ | … | $e(N_{e,t})$ |

Table 2 shows that for a given equivalent load power $u_e q$, there is always a corresponding probability $e(u_e)$. All these probabilities constitute a probabilistic sequence $e(i_{e,t})$.

### 4.2.2 Deterministic transformation of chance constraints

In order to transform the chance constraint in Eq. (16) into its deterministic form, we introduce a new type of 0-1 variable $W_{u_{e,t}}$, which satisfies the following relationship [17]:

$$W_{u_{e,t}} = \begin{cases} 1, & \sum_{n=1}^{M_G} R_{n,t}^{MT} + P_{Ress,t} \geq u_{e,t}q - E(P_t^{EL}) \\ 0, & otherwise \end{cases} \forall t, u_{e,t}=0,1,...,N_{e,t} \tag{34}$$

Eq. (34) represents that for any time period $t$, the variable $W_{u_{e,t}}$ is taken as 1 if and only if the total spinning reserve $\sum_{n=1}^{M_G} R_{n,t}^{MT} + P_{Ress,t}$ is greater than or equal to the difference of the load power $u_{e,t}q$ and the expected value $E(P_t^{EL})$; otherwise, it is taken as 0.

From Table 2, it can be observed that the probability that corresponds to the load power $u_{e,t}q$ is $e(u_{e,t})$. And thereby, Eq. (16) is simplified as

$$\sum_{u_{e,t}=0}^{N_{e,t}} W_{u_{e,t}} e(u_{e,t}) \geq \alpha \tag{35}$$

Eq. (35) suggests that during period $t$, corresponding to the all the possible output values the EL, the spinning reserve capacity in the MG meets the condition that the required confidence is greater than or equal to $\alpha$. As a result, we can now derive that Eq. (35) is equivalent to Eq. (16).

### 4.3 JAYA algorithm

JAYA algorithm is a simple yet powerful optimization algorithm proposed by R. Venkata Rao in 2016 for solving the constrained and unconstrained optimization problems. It requires only the common control parameters and does not require any algorithm-specific control parameters, which enhance the stability of the solution.

#### 4.3.1 Basic principles

This algorithm is based on the concept that the solution obtained for a given problem should move towards the best solution and should avoid the worst solution [44]. If $X_{j,k,i}$ is the value of the $j$-th variable for the $k$-th candidate during the $i$-th iteration, then the value is modified as

$$X'_{j,k,i} = X_{j,k,i} + r_{1,j,i}(X_{j,best,i} - |X_{j,k,i}|) \times r_{2,j,i}(X_{j,worst,i} - |X_{j,k,i}|) \qquad (36)$$

where $X_{j,best,i}$ ($/X_{j,worst,i}$) is the value of the variable $j$ for the best (/worst) candidate. $X'_{j,k,i}$ is the updated value of $X_{j,k,i}$ and $r_{1,j,i}$ and $r_{2,j,i}$ are the two random numbers for the $j$-th variable during the $i$-th iteration in the range [0, 1]. $X'_{j,k,i}$ is accepted if it gives a better function value.

#### 4.3.2 Hybrid coding

Considering the characteristics of controlled variables, a hybrid real/integer-coded scheme is adopted to facilitate the optimization [45]. The continuous variables comprise the $n$th ($n \in M_G$) MT's active power outputs $P_n^{MT}$, the reserve capacity $R_n^{MT}$, the reserve capacity provided by BSS $P_{B,Ress}$, the amount of trading electricity $S_B$, while the discrete variables comprise the state variable $U_n$ and start-up variable $S_n$, the BSS state variable $U_B$, the BSS charge-discharge power $P_B^{CH}$ and $P_B^{DC}$. By this means, the initial population is generated uniformly at random throughout the entire feasible search space.

### 4.4 Branch and bound algorithm

As a well-known method for discrete and combinatorial optimization problems, branch and bound algorithm (BBA) was initially proposed by A.H. Land and A.G. Doig in 1960 [46]. The main steps of BBA for minimizing an arbitrary objective function $F$ is listed as follows [47].

Step 1: Using a heuristic, find a solution $X'$ to the problem. Save the value $Y = F(X')$. $X'$ will be marked as the best solution found so far, and will be used as an upper bound on candidate solutions.

Step 2: Initialize a queue to hold a partial solution with none of the variables of the problem assigned.

Step 3: Loop until the queue is empty:

(1) Take a node $NN$ from the queue.

(2) If $NN$ stands for a single candidate solution $X$ and $F(X) < Y$, then $X$ is the best solution so far. Record it and set $Y \leftarrow F(X)$.

(3) Otherwise, branch on $NN$ to produce new nodes $NN_i$. For each of these: if $F(NN_i) > Y$, do nothing; otherwise, store $NN_i$ on the queue.

### 4.5 Determination of optimal scheme

Through solving the bi-level model, a set of scheduling schemes are obtained. To choose the optimal solution from the multiple schemes, a joint optimization objective function $F^{JO}$ is defined as follows:

$$F^{JO} = \min\sqrt{\left(F_1^{JO} - F_1^{IO}\right)^2 + \left(F_2^{JO} - F_2^{IO}\right)^2} \qquad (37)$$

where $F_1^{IO}$ denotes the IMG operating cost without consideration of the BSS, $F_2^{IO}$ is the profit of the BSS without consideration of the IMG, while $F_1^{JO}$ and $F_2^{JO}$ are respectively the net cost of the IMG and the profit of the BSS under joint optimization mode.

### 4.6 Solving process

As illustrated in Fig. 2, the solving process of the proposed approach mainly includes the follows steps:

Step 1: Model the upper IMG according to (11) ~ (16).
Step 2: Transform the chance constraints into the deterministic ones.
Step 3: Input the parameters of IMG.
Step 4: Enter an initial power price.
Step 5: Obtain a real-time price via the real-time pricing mechanism.
Step 6: Optimize the IMG model.
Step 7: Check the existence of a solution. If a solution is found, continue the optimization process; otherwise, update the confidence level and load, and then return to Step 3.
Step 8: Obtain the optimal scheduling of IMG.
Step 9: Model the lower BSS according to (17) ~ (27).
Step 10: Enter the parameters of BSS.
Step 11: Optimize the BSS model according to the real-time price via BBA.
Step 12: Obtain the charge-discharge scheduling of the BSS.
Step 13: Calculate the joint optimization objective functions $F_1^{IO}$ and $F_2^{IO}$ at the current iteration.
Step14: Termination criteria. If the predefined maximum number of iterations is met, terminate the iteration process; otherwise, return to Step 5.
Step15: Calculate the IMG operating cost $F_1^{IO}$ and the BSS profit $F_2^{IO}$ when they run independently of each other.
Step16: Determination of optimal scheme. According to (37), the minimum value of $F^{JO}$ is chosen as the optimal solution.
Step17: Output the optimal scheduling of the upper model IMG and lower model BSS, simultaneously.

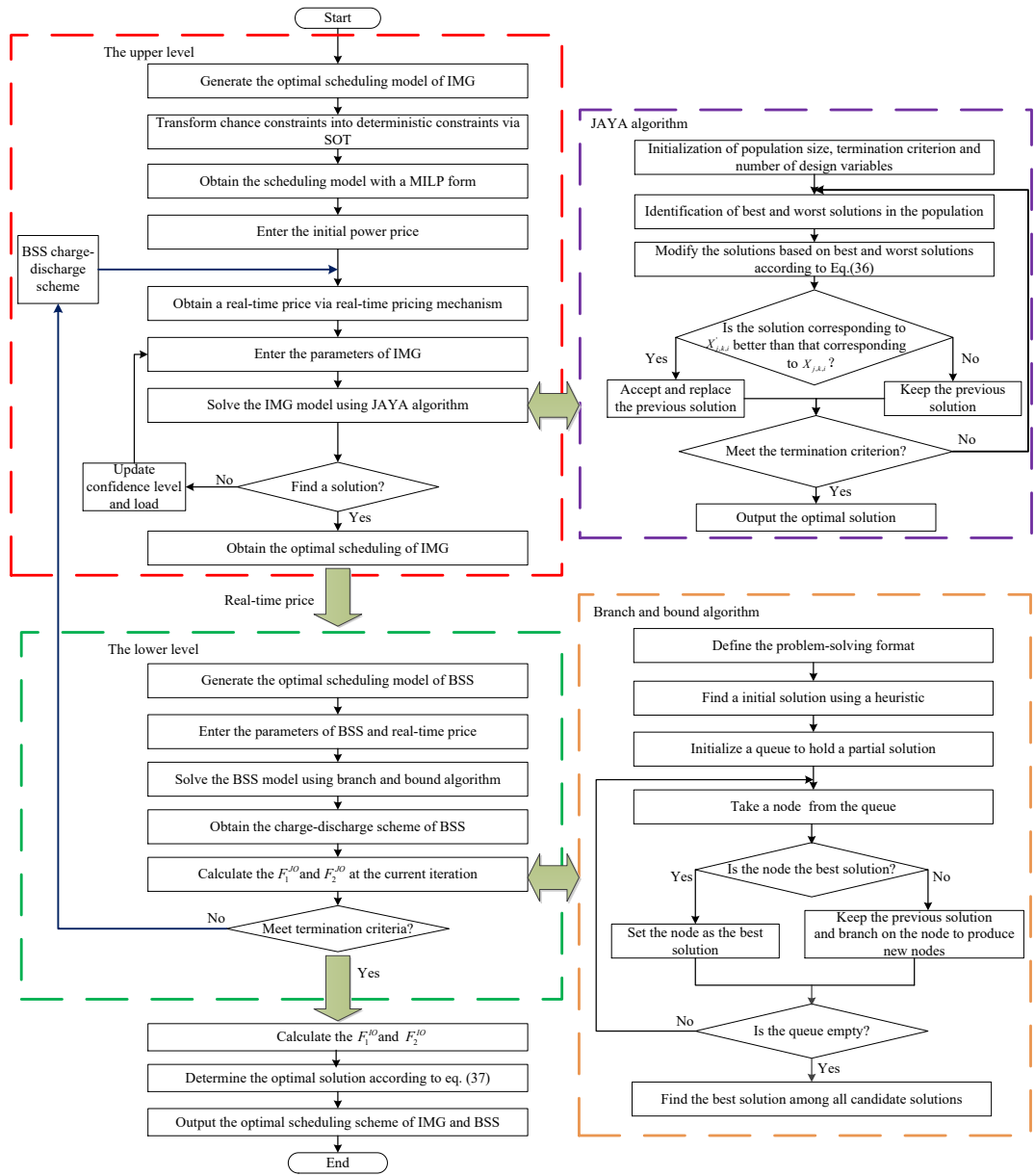

**Fig. 2.** Flowchart of the proposed solution methodology

## 5. Case study

The presented approach is tested on a modified Oak Ridge National Laboratory (ORNL) Distributed Energy Control and Communication (DECC) lab microgrid test system [15]. All the simulations are implemented on a PC platform with 2 Intel Core dual-core CPUs (2.4 GHz) and 6 GB RAM.

### 5.1 Introduction of the test system

As shown in Fig. 3, the system comprises multiple DERs, including a WT unit, a PV panel, three MT units and a BSS system, where PCC denotes the point of common coupling.

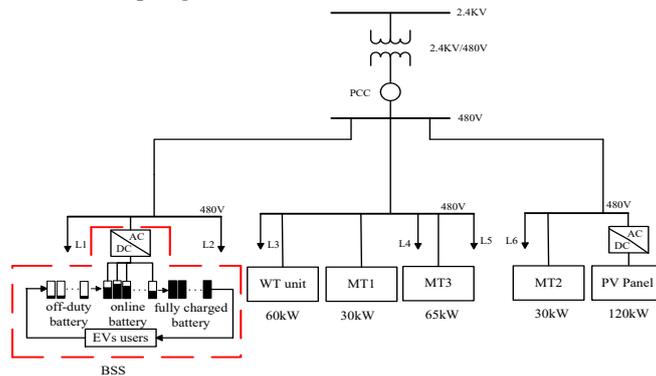

**Fig. 3.** Modified ORNL DECC microgrid test system

The parameters of the MT units are listed in Table 3.

**Table 3** Parameters of MT units

| MT number | $\zeta$ (\$) | $\kappa_n$ (\$) | $\Psi$ (\$/kW) | $\varsigma$ (\$/kW) | $P_{min}^{MT}$ (kW) | $P_{max}^{MT}$ (kW) |
|---|---|---|---|---|---|---|
| MT1 | 1.2 | 1.6 | 0.35 | 0.04 | 5 | 35 |
| MT2 | 1.2 | 1.6 | 0.35 | 0.04 | 5 | 30 |
| MT3 | 1.0 | 3.5 | 0.26 | 0.04 | 10 | 65 |

The parameters of the WT are as follows: $v_{in}=3$m/s, $v_r=15$m/s, $v_{out}=25$m/s, and $P_r=60$kW, $P_{min}^{WT}$ and $P_{max}^{WT}$ are the minimal and maximum value of the WT power outputs throughout the entire scheduling period. The parameters of the PV panel are given as follows: $\eta^{PV}=0.093$, $A^{PV}=1300\text{m}^2$, and $P_{max}^{PV}=120$kW. The maximum value of the load power is $P_{max}^L=195$kW. The parameters of the BSS are as follows: $P_{max}^{CH}=7.5$kW, $P_{max}^{DC}=15$kW, $C_{bat,max}=19$kW·h, $\eta_B^{DC}=\eta_B^{CH}=0.95$, $\tau=3$ \$ [48], $\omega_{sp\_price}=1.4$ \$/kWh, $\omega_{rc\_price}=0.02$ \$/kW, $\omega_{rf\_price}=1.0$ \$/kWh, $P_{rf}^{EL}=80$ kW, $N_{bat,cyc,max}=2$ cycles. In this study, there are four charge-discharge positions and sixteen leasable batteries in BSS.

## 5.2 Simulation scenarios

### 5.2.1 Outputs of DGs and load

Based on the above-mentioned probabilistic models in Section 2 and the related parameters in Section 5.1, the expected value of DG outputs and load power during all time periods can be obtained, and thereby that of the equivalent load can be figured out. These data will be employed as the basic data for the follow-up analysis, as illustrated in Fig. 4.

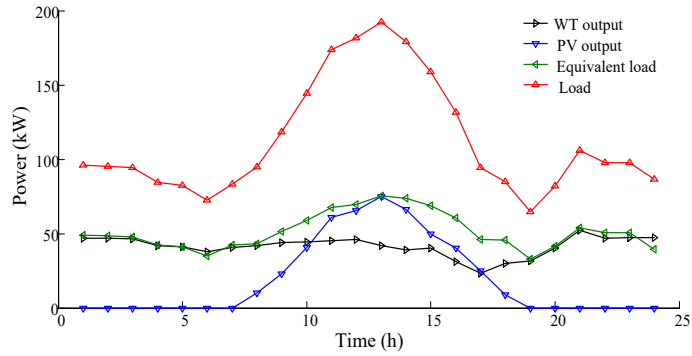

**Fig. 4.** The power curves of distributed generation and load

From Fig. 4, we can see that based on the statistical characteristics of WT, PV and load outputs, the SOT is capable of dealing with the superposition of multiple uncertainties, which further confirms the conclusions of the work in reference [36].

### 5.2.2. Swap demand of EVs

As mentioned above, we postulate that the number of EVs arriving at the BSS during a period follows the Poisson distribution, and then the swap demand of EVs during a typical scheduling cycle (a day) is obtained as shown in Fig. 5.

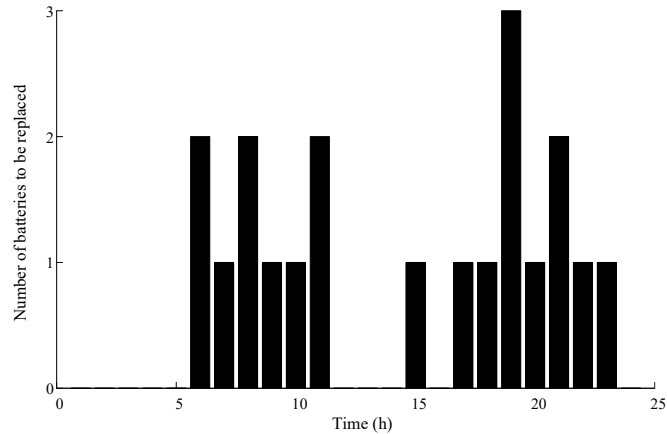

**Fig. 5.** The swap demand of EVs during a scheduling cycle

From Fig. 5, it can be observed that the swap demand of EVs has an obvious randomness. As mentioned earlier, from the perspective of the MG, BSS can be considered as an uncertain load demand. Therefore, the swap demand further exacerbates the uncertainty of uncertainty of MG operation.

## 5.3 Economic costs analysis

In order to explain clearly the interest advantage of joint optimization between IMG and BSS, three different optimization strategies are discussed in this section. More specifically, IMG independent optimization, joint optimization, and BSS independent optimization are respectively named as strategy 1, 2 and 3. These studies were conducted at a discrete step of 2.5 kW, a confidence level of 90 %, and a load fluctuation of 10 %. Note that the IMG net costs in strategies 1 and 3 are equal to the operating costs of the IMG in subsequent analysis since the microgrid does not generate any revenue in such cases.

### 5.3.1. Independent optimization analysis of IMG and BSS

The optimal objective function value of IMG independent optimization is obtained according to the JAYA optimization algorithm. According to the adjustable power of each time period provided by IMG, BBA is used to carry out independent optimal dispatching of BSS. As shown in Fig. 3, a bi-directional converter connecting BSS and IMG is utilized to control the power flow for their power exchange in this work. To facilitate analysis, we assume that the BSS charge-discharge power maintains a constant power, and there is no power loss when the power flows through the bi-directional converter. The exchanging powers between BSS and IMG under three strategies are shown in Fig. 6, where a positive/negative power respectively denotes the charge/discharge power of the BSS.

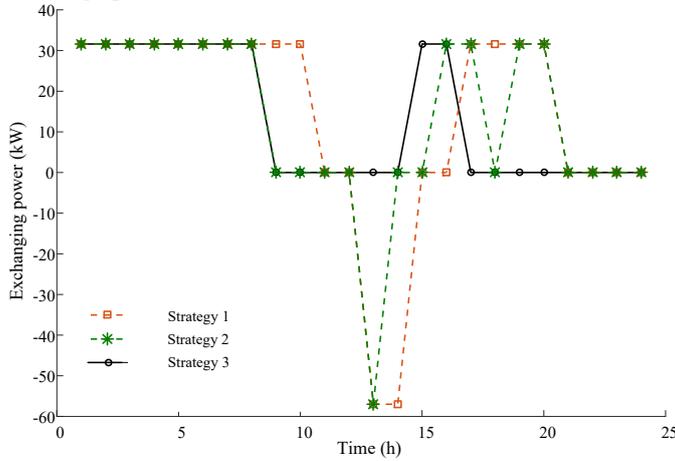

**Fig. 6.** The exchanging power under different optimization strategies

Fig. 6 shows that the BSS plays different roles under different load conditions under strategy 2 (joint optimization). Compared with the strategy 1, the BSS, as a power source, sells electricity to the IMG during the peak load period, which can improve the profits of the BSS; while compared with the strategy 3, the BSS, as a power load, absorbs surplus electricity of the IMG during the valley load period, which can reduce the net costs of the IMG.

### 5.3.2. Joint optimization analysis

JAYA-BBA algorithm is used to solve the joint optimal scheduling strategy in this paper. The input parameters of JAYA algorithm include maximum generation and population size that are set to 1500 and 100 respectively in this paper. The optimization results are shown in Fig. 7.

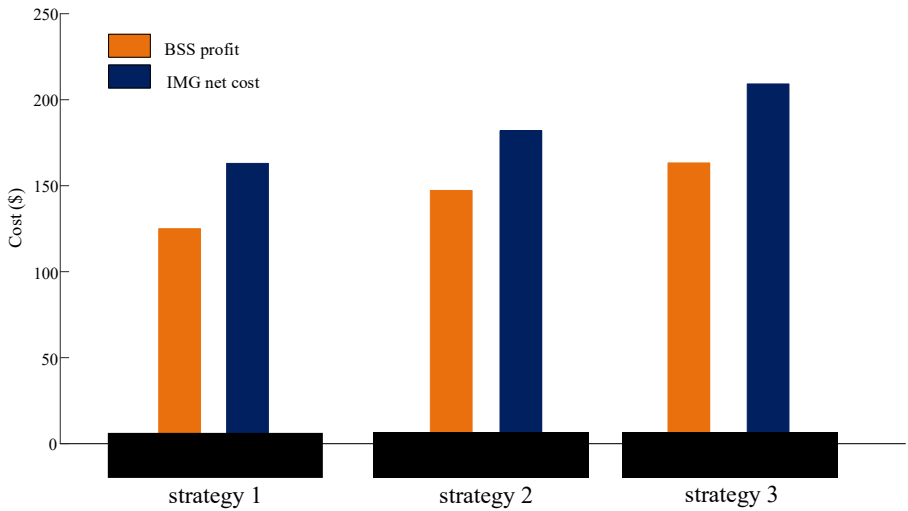

**Fig. 7.** The BSS profits and IMG net costs under different optimization strategies

Fig.7 suggests that the optimization results are significantly different under different strategies. Concretely speaking, the net cost of the IMG is the lowest under strategy 1, but it also leads to the least BSS profit; the BSS profit reaches the optimal solution under strategy 3, but the corresponding IMG net cost is the highest. Compared with strategy 3, the net cost of the IMG is reduced under strategy 2, while at the same time satisfying the economic operation of the BSS. This result indicates that a balance between IMG net cost and BSS profit can be achieved by coordinating the scheduling between IMG and BSS.

The hourly variations in the IMG net costs and the BSS profits during an entire scheduling cycle (i.e, a whole day) are shown in Fig. 8.

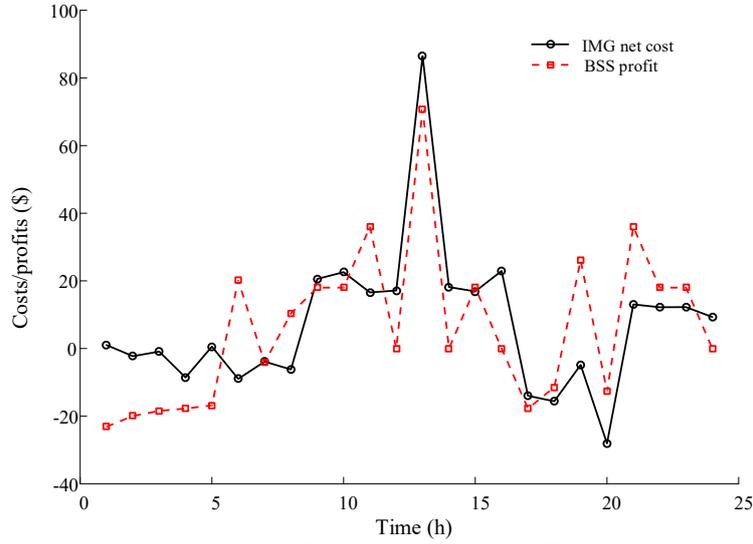
**Fig. 8.** The economy of IMG and BSS during different periods

From Fig 8, it can be seen that the IMG net cost and BSS profit have distinctly different characteristics. Regarding the IMG, its net cost during the BSS charging period (00:00-08:00, 16:00-20:00) is significantly lower than that of other periods. The main reason for this is that the BSS purchases power from the IMG, which results in the decrease in the IMG net cost due to the offset of the profits from selling electricity to the BSS. As far as the BSS is concerned, during most of the periods that the BSS provides power swapping services, the BSS profits are obviously greater than other periods, since the swapping power services increase the BSS profits. In addition, we can also observe that both the IMG net cost and the BSS profit reach their maximum value during 12:00-13:00 period due to the fact that the IMG purchases electricity from the BSS during this period to meet the balance between supply and demand of the microgrid system.

Taking interval 18:00-19:00 as an example, a comparative analysis of scheduling results under different strategies is shown in Table 4. Note that a negative value of IMG net cost indicates that the IMG is in a profitable state.

**Table 4** Optimization results analysis of BSS and IMG at 19:00

| Parameter | BSS independent optimization | Joint optimization | IMG independent optimization |
|---|---|---|---|
| WT output (kW) | 31.75 | 31.75 | 31.75 |
| PV output (kW) | 0 | 0 | 0 |
| Load (kW) | 64.65 | 64.65 | 64.65 |
| EL (kW) | 32.9 | 32.9 | 32.9 |
| MT3 output (kW) | 32.9 | 64.5 | 64.5 |
| MT2 output (kW) | 0 | 0 | 0 |
| Exchanging power (kW) | 0 | 31.6 | 31.6 |
| MT1 reserve capacity (kW) | 0 | 30.37 | 30.37 |
| BSS reserve capacity (kW) | 30.37 | 0 | 0 |
| Price ($) | 1.269 | 0.85 | 1.269 |
| BSS profit ($) | 54 | 26.14 | 13.90 |
| IMG net cost ($) | 14.56 | -4.84 | -18.08 |

As shown in Table 4, the optimization results of different operating modes are different from each other. Specifically, the BSS independent optimization produces the highest BSS profit and the highest IMG operating cost, while the IMG independent optimization yields the lowest BSS profit and the highest IMG operating profit. Moreover, it can be observed that the result of the joint optimization strategy is a trade-off between those of the former two strategies.

The reasons are analyzed as follows. (1) Compared with BSS independent optimization, the IMG net cost of the joint optimization decreases and switch to a profitable state because of two factors. On one hand, the fuel costs of MT units increase due to the increase of MT outputs, and at the same time, the reserve costs increase as well since the shortage of reserve capacity has to be provided by MT units, while the reserve price of MTs is greater than that of the BSS. But on the other hand, the profits of the IMG resulting from the charging of the BSS offset the increase of IMG operating cost. According to Eq. (11), it is because the profit of IMG is greater than the increase in its costs in this period that the IMG is profitable on the whole. (2) Compared with IMG independent optimization, the BSS yields higher profits under the joint optimization mode. According to Eq. (17), we know that the BSS profits are affected by its charge-discharge prices. Under the mode of joint optimization, real-time prices, rather than grid prices, are utilized as the BSS charge-discharge prices, which result in higher BSS profits for the same amount of trading electrical quantity. Therefore, the conclusion that can be drawn on the basis of the above facts is that the proposed approach manages to coordinate the scheduling problem between IMG and BSS with the use of bi-level programming.

## 5.4 Demand respond analysis

In order to verify the effectiveness of the price-based demand response mechanism, two cases are designed and analyzed.

Case 1: The grid price is used as the exchanging energy price between IMG and BSS without consideration of demand responds.

Case 2: The real-time price is used as the exchanging energy price between IMG and BSS with consideration of demand responds.

### 5.4.1 Charge-discharge scheduling

In Case1, the BSS charge-discharge scheduling is shown in Fig. 9.

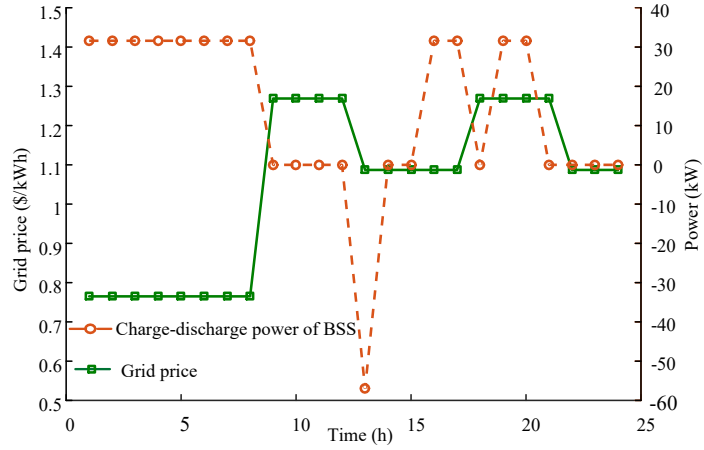

**Fig. 9.** The charge-discharge power curves of BSS in case1

It can be seen in Fig. 9 that it is inappropriate to use the grid price as the trading price between IMG and BSS in this case, since their operating characteristics are not fully taken into consideration. More specifically, though the trading price is relatively low during period 12:00-13:00, the BSS has to sell electric power to the IMG for the purpose of balancing supply and demand in the IMG system; while during period 18:00-20:00 the BSS has no choice but to purchase electricity from IMG at a relatively high price period to satisfy the swap demand of EVs. From the view of BSS, this unreasonable charge-discharge plan is contrary to market trading principles that the BSS should discharge power at a high price and charge at a low price. As far as the IMG, there is a heavy load demand during period 15:00-16:00 as shown in Fig. 4, but the BSS absorbs electric energy, which is not conducive for the IMG to maintain its supply and demand balance.

In Case 2, the charge-discharge scheduling for the BSS is shown in Fig. 10.

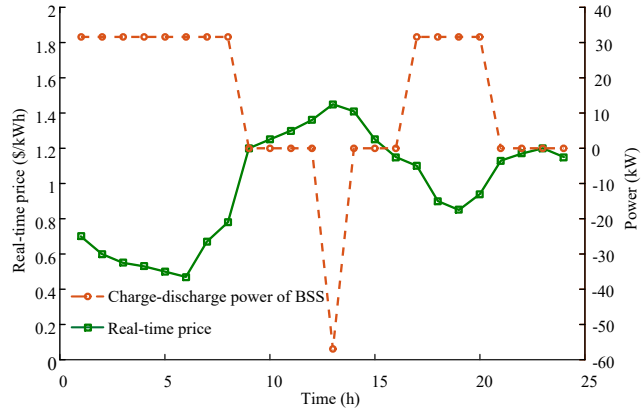

**Fig. 10.** The charge-discharge power curves of BSS in Case 2

From Fig. 10, it can be observed that the BSS charge-discharge activities in Case 2 are obviously different from those in Case 1. In Case 2, the BSS is charged during the low-price periods to meet the swap demand for EVs users; while during the high-price periods, the BSS sells electricity to the IMG to keep the IMG's supply-demand balance. Furthermore, compared with Case 1, there will be no such situations in which the BSS sells electricity to the IMG during the low-price periods and purchases electricity from the IMG during the high-price periods in Case 2. As a result, a conclusion can be safely drawn based on this fact that the real-time pricing based on demand responses is capable of promoting the active participation of the BSS in the regulation of the IMG operation to achieve win-win outcomes.

#### 5.4.2 MT operation analysis

For a given confidence level such as 90%, MT3 and MT2 play a role of outputting powers to balance between supply and demand of the system, while MT1 is only responsible for providing spinning reserve services to the system. The operation curves of MT2 and MT3 in the two cases are shown in Fig. 11.

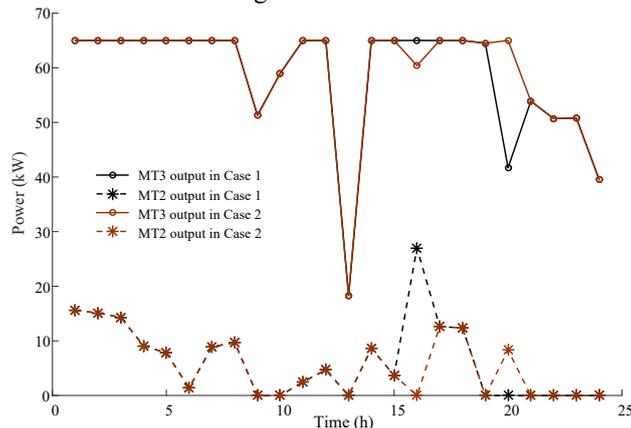

**Fig. 11.** The outputs of the MTs in different cases

Fig. 11 shows that in both cases, the outputs of MT3 are always greater than those of MT2. In fact, only when MT3 are unable to balance the difference between supply and demand does MT2 start to provide power outputs. The reason for this phenomenon is that the consumption coefficients of MT3 are less than those of MT2. As a result, MT3, rather than MT2, services a primary power source to achieve a lower operating cost of the system.

And at the same time, it can also be observed in Fig. 11 that the MTs' outputs in Case 2 are smoother than those in Case 1, which is specifically reflected in the two intervals, i.e. 15:00-16:00 and 19:00-20:00. The reason for this is that the MT outputs are subject to the BSS charge-discharge scheme determined by the battery swapping demand of EVs and charge-discharge prices. Therefore, the real-time pricing mechanism based on demand responses play an important role in peak-load shaving and suppress the fluctuation of MTs output.

### 5.5 Influence analysis under different confidence levels

#### 5.5.1 BSS operation schedule analysis under different confidence levels

In order to analyze the effects of different confidence levels on BSS operation schedules, the BSS charge-discharge powers under different confidence levels during a scheduling cycle are shown in Fig. 12.

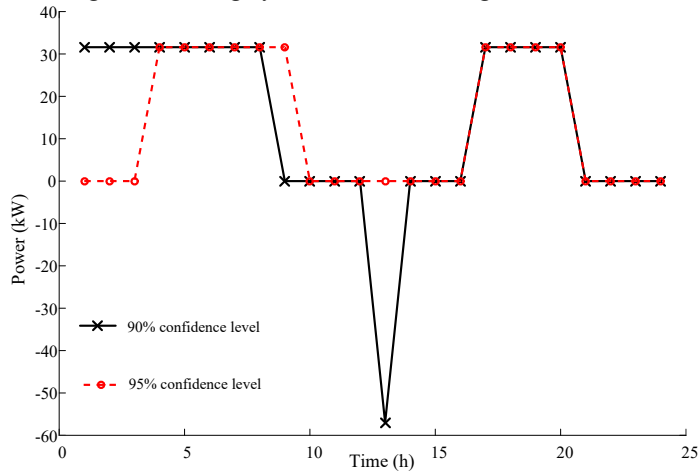

**Fig. 12.** The charge-discharge schedule of BSS under different confidence levels

Fig. 12 shows that at the 90% confidence level, the BSS discharges to the IMG; while at the 95% confidence level, the BSS does not do it. This can be explained as the available capacity of the BSS significantly decreases at a higher confidence level since the IMG requires more spinning reserves to ensure its operational reliability; while on the other hand, regarding the BSS, a higher priority should be given to satisfying the swap demand of EVs, rather than supplying power to the IMG. As a result, the BSS does not discharge to the IMG at the 95% confidence level due to the insufficient remaining capacity.

#### 5.5.2 Reserve capacity analysis under different confidence levels

In order to analyze the spinning reserves under different confidence levels, the total reserve capacity required by the system and its supply situations are respectively shown in Figs. 13 and 14.

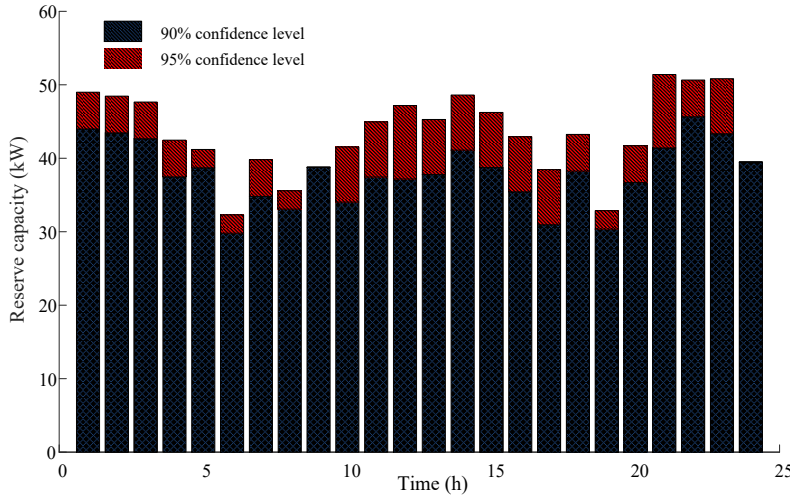

**Fig. 13.** Total reserve capacities under different confidence levels

As shown in Fig. 13, during the period 7:00-18:00, the confidence level can reach 90% when the reserve capacity maintains in the range between 35 and 40 kW; while during the period that the WT works alone, more spinning reserve capacity needs to be prepared to maintain the 90% confidence level. This demonstrates that the complementary nature of different types of DERs can reduce the random fluctuations in the DG outputs. On the other hand, it can be seen from Fig. 13 that with the increased confidence level, the system needs to configure more spinning reserve capacity which inevitably increases the operating cost. Therefore, it is substantial to select appropriate confidence levels to achieve a better balance between reliability and economy.

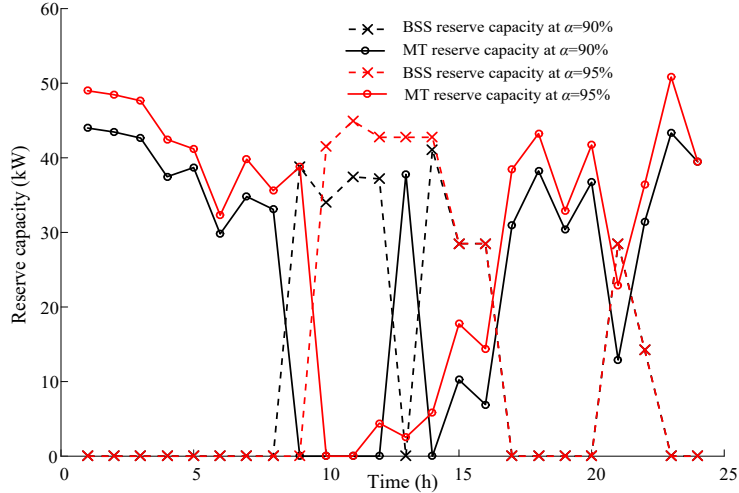
**Fig. 14.** Respective reserve capacities provided by the MTs and BSS under different confidence levels

From Fig. 14, it can also be seen that with the increase in confidence levels, both BSS and MTs provide more reserve capacities to ensure that the system operates under the given reliability. More importantly, it can be seen that selection of a reasonable confidence level is critical for the system operation. (1) During period 0:00-8:00, BSS is only charged from IMG as a load, and all the needed reserve capacities are provided by the MTs. (2) During period 8:00-16:00, the reserve capacity provided by the BSS is basically greater than that from the MTs under each confidence level. The reason for this is that the BSS take precedence over MTs to provide reserves due to its lower costs and faster response speed. Only if the remaining capacity of the BSS is not enough to bear the required reserves, the MTs are employed to provide reserves. (3) During period 16:00-20:00, all the reserves are provided by the MTs, since the batteries in the BSS are in the state of charge at this moment. (4) During period 20:00-24:00, the BSS provides reserve capacities for the IMG and swap services for EV users at first, and then the increasing reserves have to be afforded by the MTs with the gradual decrease of battery remaining energy resulting from EV swap services.

### 5.6 Economic comparison

In order to reasonably evaluate the effects of real-time pricing on the different stakeholders (IMG and BSS), an economic comparison with non-considerations of demand responses has been made, and the IMG net costs and the BSS profits in different cases are listed in Table 5.

**Table 5** IMG net costs and BSS profits

| | IMG | BSS | | | | |
|---|---|---|---|---|---|---|
| Case | Net costs ($) | Discharging incomes ($) | Swapping incomes ($) | Charging costs ($) | Depreciation costs ($) | Total profits ($) |
| 1 | 182.19 | 61.96 | 420 | 280.33 | 72 | 139.66 |
| 2 | 176.43 | 82.65 | 420 | 272.85 | 72 | 147.15 |

Table 5 shows that compared with Case 1, the economics of both the IMG and BSS have been improved in Case 2 due to the introduction of the real-time pricing mechanism. Specifically, the IMG net cost has been reduced by 3.16%, while the BSS profit has been increased by 5.36%. In this way, the different stakeholders achieve a win-win situation. Therefore, the effectiveness of the proposed real-time pricing mechanism is verified.

### 5.7 Comparison with other methods

In order to assess the performance of the proposed method, a commonly used hybrid intelligent algorithm (HIA) approach which combines the particle swarm optimization (PSO) algorithm with Monte Carlo simulations (MCS) [49-52] is utilized as a comparison algorithm for solving the model. The specific parameters of PSO and MCS are set as follows. For PSO, the population size is set to 20, and the maximum number of iterations is 150; for MCS, the number of random variables $N_s$ is 500. Due to the uncertainties of the HIA [15], the results of the HIA shown here are the average values of 20 runs.

Let $\sigma_L$ =10%, and the step size $q$ is taken as 2.5 kW. The comparison results are shown in Table 6.

**Table 6** Comparison results

| Confidence levels (%) | Proposed approach | | | HIA | | |
|---|---|---|---|---|---|---|
| | Economy ($) | | Calculation time (s) | Economy ($) | | Calculation time (s) |
| | IMG cost | BSS profit | | IMG cost | BSS profit | |
| 80 | 166.44 | 139.09 | 36.7 | 173.29 | 133.37 | 372.6 |
| 85 | 171.04 | 143.69 | 38.3 | 178.31 | 137.16 | 347.1 |
| 90 | 176.43 | 147.15 | 37.5 | 183.16 | 140.23 | 364.7 |

From Table 6, it can be found that the results of the proposed approach are better than those of the HIA method in the following two aspects: first, the obtained net cost of IMG and the profit of BSS using our approach are respectively superior to the results of the HIA under all different confidence levels; second, the computation times of the proposed method are less than 40 seconds, which are remarkably less than those of the HIA. It can be expected that if more advanced hardware configurations are used, the computational efficiency will be further improved. Therefore, the conclusion can be drawn that the Jaya-BBA is an effective tool to address the bi-level programming problem, and that the computational efficiency of the proposed method meets the real-time requirements of the microgrid scheduling. Furthermore, the superiority of our approach relative to other the advanced HIA method has also been verified, which reflects in reduced costs, greater revenues and less calculation time.

## 6. Conclusion

Confronting the increasingly serious energy crisis and environmental pollution problems, research on MG and EVs has attracted ever-growing concerns in recent years. To alleviate this dilemma, it's a 'win-win' strategy to integrate BSSs into an MG for both of them. Considering their inherent hierarchical features, the optimal day-ahead scheduling of IMG with a BSS in multi-stakeholder scenarios has been investigated in this work using a bi-level programming approach, in which a real-time pricing mechanism based on demand responses is designed to reflect the dynamic supply-demand relationships between IMG and BSS. To solve the bi-level programming model, a hybrid approach combining the heuristic with analytical optimization, called JAYA-BBA, is developed. And finally, simulation experiments have been carried out on the modified ORNL DECC lab MG test system to examine the effectiveness of the proposed approach. The simulation results demonstrate that the proposed model can effectively promote the participation of BSSs in regulating the IMG economic operation, and furthermore, the superiority of our approach to the state-of-the-art HIA method is verified as well, embodying in the better economic benefits and faster solution efficiency.

Our future work will focus on the multi-timescale scheduling by integrating day-ahead and real-time scheduling sub-models. It is interesting to investigate the prediction sequence model of the electric vehicle arriving by using a machine learning technique according to a great deal of historical data. In this paper, the lifetime of batteries can be prolonged to a certain extent by limiting the number of charge/discharge cycles, while, to maximize the battery life cycle, one needs to calculate the cycle life of the battery during the optimization problem [43]. Besides, more realistic modeling techniques will be developed to deal with the correlations between load and DER uncertainties [17]. Another interesting topic for further research is to extend this study to potential applications in the optimal operation of a micro integrated energy system [53].

**Acknowledgements**

This work is supported by the U.S. Department of Energy (DOE)'s Office of Electricity Delivery and Energy Reliability – Advanced Grid Modeling (AGM) Program and the China Scholarship Council (CSC) under Grant No. 201608220144.